\documentclass[
 reprint,
 amsmath,amssymb,
 aps,
 superscriptaddress,
 nofootinbib,
]{revtex4-2}

 \usepackage[skins,theorems]{tcolorbox}
\tcbset{highlight math style={enhanced,
  colframe=red,colback=white,arc=0pt,boxrule=1pt}}
  \usepackage[bookmarksopen, bookmarksnumbered, bookmarksopenlevel=2]{hyperref}
  \usepackage{tikz}
  \usepackage{tikz-3dplot}
  \usepackage{float}
 \usetikzlibrary{calc}
 \usetikzlibrary{decorations}
 \usepackage[UKenglish]{babel}
 \usepackage[toc,page]{appendix}
 \usepackage{amsmath}
 \usepackage{amssymb}
 \usepackage{graphicx}
 \usepackage{hhline}
 \usepackage[bf]{caption}
 \usepackage{xcolor}
\usepackage[vcentermath]{youngtab}
\usepackage{geometry}
\usepackage{slashed}
\usepackage{color}
\usepackage{stackrel}
\usepackage{tikz-cd} 
\usepackage{mathtools}
\usepackage{cancel} 
\usepackage{multirow}
\usepackage[margin=0pt,font=small,labelfont=normalfont,skip=22pt]{subcaption}
\usepackage{empheq}
\usepackage{color,soul}
\usepackage{arydshln}
\usepackage{physics}
\usepackage{amsmath}
\usepackage{tikz}
\usepackage{mathdots}
\usepackage{yhmath}
\usepackage{cancel}
\usepackage{color}
\usepackage{siunitx}
\usepackage{array}
\usepackage{float}
\usepackage{multirow}
\usepackage{amssymb}
\usepackage{gensymb}
\usepackage{tabularx}
\usepackage{extarrows}
\usepackage{booktabs}
\usetikzlibrary{fadings}
\usetikzlibrary{patterns}
\usetikzlibrary{shadows.blur}
\usetikzlibrary{shapes}

\newcommand{\bqa}{\begin{eqnarray}}
\newcommand{\eqa}{\end{eqnarray}}

\newenvironment{eqn*}{\begin{equation*}\begin{aligned}}{\end{aligned}\end{equation*}\noindent}
\hypersetup{
    pdftitle={},
    pdfauthor={},
    pdfsubject={}
}
\makeatletter

\newcommand{\be}{\begin{equation}}
\newcommand{\ee}{\end{equation}}
\newcommand{\beq}{\begin{equation}}
\newcommand{\eeq}{\end{equation}}
\newcommand{\ba}{\begin{aligned}}
\newcommand{\ea}{\end{aligned}}

\newcommand{\bea}{\begin{eqnarray}}
\newcommand{\eea}{\end{eqnarray}}

\newcommand{\cO}{\mathcal{O}}

\newcommand{\cN}{\mathcal{N}}

\newcommand{\cR}{\mathcal{R}}

\newcommand{\cM}{\mathcal M}

\newcommand\bi{\begin{itemize}}
\newcommand\ei{\end{itemize}}

\def\unit{{1\kern-.65ex {\rm l}}}
\def\1{{1\kern-.65ex {\rm l}}}

\def\dd{{\mathrm{d}}}
\def\pl{{\mathrm{pl}}}

\newcount\hour \newcount\minute
\hour=\time \divide \hour by 60
\minute=\time
\def\now{%
\ifnum \hour<13
  \ifnum \hour=0 \advance \hour by 12 \number\hour:\else \number\hour:\fi%
     \ifnum \minute<10 0\fi%
     \number\minute%
\ A.M.%
\else \advance \hour by -12 \number\hour:%
  \ifnum \minute<10 0\fi%
  \number\minute%
  \ P.M.%
\fi%
}

\def\p@subsection{}
\def\p@subsubsection{}

\makeatother

\begin{document}

\title{The Tale of Three Scales:\\the Planck, the Species, and the Black Hole Scales}

\author{Alek Bedroya}
\affiliation{Princeton Gravity Initiative, Princeton University, Princeton, NJ 08544, USA}
\affiliation{Jefferson Physical Laboratory, Harvard University, Cambridge, MA 02138, USA}

\author{Cumrun Vafa}
\affiliation{Jefferson Physical Laboratory, Harvard University, Cambridge, MA 02138, USA}

\author{David H. Wu}
\affiliation{Jefferson Physical Laboratory, Harvard University, Cambridge, MA 02138, USA}

% \date{\today}

\begin{abstract}
Quantum gravity (QG) has a natural cutoff given by the Planck scale $M_{\pl}$.   However, it is known that the EFT of gravity can break down at a lower scale, the species scale $\Lambda_s\lesssim M_{\pl}$, if there are light species of particles.  Here we point out that there is a third scale $\Lambda_{\rm BH}\lesssim \Lambda_s\lesssim M_{\pl}$, which marks the inverse length (or the temperature) of the smallest black hole where the EFT gives a correct description of its entropy and free energy.  This latter scale is hard to detect from the viewpoint of EFT as it represents a phase transition to a state with lower free energy.  We illustrate this using examples drawn from consistent QG landscape.  In particular $\Lambda_{\rm BH}$ gets related to Gregory--Laflamme transition in the decompactification limits of quantum gravity and to the Horowitz--Polchinski solution in the light perturbative string limits.  We propose the existence of $\Lambda_{\rm BH}$ marking the temperature at which neutral black holes undergo a phase transition, as a new Swampland condition for all consistent quantum theories of gravity.  In the asymptotic regimes of field space $\Lambda_{\rm BH}$ is close to the mass scale of the lightest tower but deviates from it as we move inwards in the moduli space.

\end{abstract}

\maketitle

\setcounter{page}{1}
\section{Introduction}
Quantum gravitational theories enjoy a natural cutoff, the Planck scale $M_{\pl}$.  When curvature is of the order of $M_{\pl}^2$ the Einstein term in the action becomes strong and the quantum fluctuations render the classical picture of gravity inadequate.  However, the actual scale in which the quantum gravity effective action can become invalid may be smaller.  Indeed, it was argued in \cite{Dvali:2007hz,Dvali:2009ks,Dvali:2010vm,Dvali:2012uq} that if there are light species of particles present in the theory, the consistency of black hole entropy with the \textit{effective field theory} (EFT) description requires the breakdown of the classical picture at a lower scale called the ``species scale'', $\Lambda_s$.  It was proposed in \cite{vandeHeisteeg:2022btw} that the mechanism for this is that the higher-order corrections to the gravitational action become strong at the scale $\Lambda_s$ instead of $M_{\pl}$ as one may naively expect.  Even though the introduction of $\Lambda_s$ was motivated by the appearance of the tower of light states, surprisingly the mass scale of this tower, $m$, does not seem to be directly captured by $\Lambda_s$.  Indeed, the EFT does not directly encode the mass scale of the tower.  For example, when the spectrum of light states is dominated by KK modes coming from decompactifying a $d$ dimensional theory to $D$ dimensions, the species scale  corresponds to the Planck mass in the higher-dimensional theory $\Lambda_s^{D-2}\sim M_{\pl}^{d-2}R^{d-D}$, and not the KK mass scale $m_{\rm KK}=1/R$.  In other words, the number of extra dimensions $D-d$ is harder to detect from the low energy EFT. Note that a sufficiently high precision measurement of the gravitational amplitudes can always determine $m_{\rm KK}$. However, as explained in the Appendix~\ref{appendix}, the corrections to the EFT coefficients at energy scale $\mu\ll \Lambda_s$ that are controlled by $m_{\rm KK}$ are always sub-leading, either to the massless threshold contribution or to the terms that are powers of the species scale. One way the lower-dimensional theory can find out about this scale is in the study of black holes. When black holes get smaller than $R$, the black hole becomes thermodynamically unstable due to the Gregory--Laflamme transition \cite{Gregory:1993vy}: the larger black holes correspond to black branes wrapped around the extra dimensions, but the smaller ones are localized in the extra dimensions. This transition to a new black hole solution marks a new scale in the lower-dimensional theory, which we denote as $\Lambda_{\rm BH}$. A similar phenomenon occurs when we go to regions in field space with light fundamental strings. In this case, the species scale can be identified with the Hagedorn temperature which is proportional to the string scale $\Lambda_s\sim M_s$. However,
as we consider high-temperature black holes, there is evidence that before we get to the Hagedorn temperature, there is another scale where the black holes undergo a transition to a more stable state, the Horowitz-Polchinski solution, which we identify as $\Lambda_{\rm BH}$.  These two types of towers are the only types expected in the large-distance limits in quantum gravity \cite{Lee:2019wij}. Motivated by these observations, we propose a new Swampland condition: For all quantum gravitational theories, there is a temperature scale, $\Lambda_{\rm BH}$, which is the lowest scale at which the black hole describable by the EFT becomes unstable due to a transition into a more stable saddle:  $\Lambda_{\rm BH}\lesssim \Lambda_s\lesssim M_{\pl}$.  Moreover, this scale marks the existence of a phase transition from the black hole well described by the low-energy EFT to a more stable solution which is not visible to that EFT. We explain how this is related to the fact that massive weakly coupled states, such as towers, are omitted from the dynamical spectrum of the low-energy EFT.  In particular, we propose that in the context of type IIA string theory a transition analogous to the Horowitz--Polchinski transition continuously deforms to the Gregory--Laflamme transition as we move from weak string coupling to strong coupling.

The organization of the rest of this paper is as follows: In section~\ref{sec:The species scale}, we review the species scale. In section~\ref{sec:Phase transitions}, we review the Gregory--Laflamme transition and the Horowitz--Polchinski transition.
In section~\ref{sec:Swampland conjecture}, we conclude by proposing a new Swampland condition.

\section{The species scale $\Lambda_s$}
\label{sec:The species scale}

In the EFT approach to quantum gravity in $d\geq 4$ spacetime dimensions, the leading terms are given by the Einstein action. The EFT is defined with a natural energy scale $\mu$ to describe backgrounds with curvature scales $\mathcal{R}\sim\mu^2$. Additionally, there will be higher-order terms which are suppressed for low-curvature regions.  The EFT is expected to break down when these higher-order terms become of the same order as the leading term. 
Then, the coefficients of these higher-order terms, denoted as $a(\phi_i,\mu)$, can depend on the massless scalar fields $\phi_i$ that parameterize the moduli space $\cM_{\rm QG}$ as well as $\mu$. Hence, the higher-order corrections appearing in the action can be written as
\begin{equation}
\label{eq:supergravity}
    S_{\mathrm{corr.},d}\supset \int \dd^dx\sqrt{-g}\left[\sum_n a_n(\phi_i,\mu)\cO_{n}(\cR)\right]\,,
\end{equation}
where $\cO_{n}(\cR)$ denotes $2n$-dimensional operators.\footnote{Note that our convention for the coefficients of higher-derivative operators differs from the definition in \cite{vandeHeisteeg:2023dlw} by power of Planck mass.} The dependence of the EFT coefficients on the moduli would be more pronounced when there are a large number of light species due to loop corrections. One would expect that the leading terms should define a new moduli-dependent physical energy scale. This is the energy scale at which the sub-leading corrections $a_n(\phi_i,\mu)\mathcal{O}_n(\mathcal{R})\sim a_n(\phi_i,\mu)\mu^{2n}$ become of the same order as the Einstein term $M_{\pl,d}^{d-2}\mathcal{R}\sim M_{\pl,d}^{d-2}\mu^2$. This scale is identified as the species scale $\Lambda_s(\phi_i)$
\begin{equation}\label{SCD}
    \frac{1}{\Lambda_s(\phi_i)^{2n-2}}\sim \frac{a_n(\phi_i,\Lambda_s(\phi_i))}{M_{\pl,d}^{d-2}}\,,
\end{equation}
where the above relation is expected to hold up to constants for generic higher-order terms in the action.
In other words, whereas one naively expects that the higher-order terms become relevant when the curvature is of the order of $M_{\pl}^2$, it actually breaks down sooner at curvature scale $\Lambda_s(\phi_i)^2\lesssim M_{\pl}^2$ corresponding to energy scale $\Lambda_s(\phi_i)$. Hence, $\Lambda_s(\phi_i)$ can be viewed as an effective gravitational UV cutoff. An interesting feature of this definition is that the dependence of species scale across the full moduli space can in some situations be exactly calculated using supersymmetrically protected terms in the action \cite{vandeHeisteeg:2022btw,vandeHeisteeg:2023dlw}!

To ensure the definition \eqref{SCD} of $\Lambda_s$ correctly captures the strong-coupling behavior of gravity, one can test it in top-down constructions in string theory. In the corners of the moduli space with a weakly coupled string, we expect the effects of quantum gravity to appear at string mass, i.e., $\Lambda_s\sim M_s\sim \exp\left(-\phi/\sqrt{d-2}\right)$ where $-\phi$ is the canonically normalized dilaton in the reduced Planck units. Furthermore, in M-theory corners where there is no weakly coupled string, the quantum gravity effects are expected to be controlled by a higher D-dimensional Planck mass, i.e., $\Lambda_s\sim M_{\pl,D}\sim \exp\left(-\phi \sqrt{\frac{D-d}{(D-2)(d-2)}}\right)$ where $-\phi$ here is the canonically normalized volume modulus in the reduced Planck units. It has been conjectured that these two examples cover all weak-coupling limits in quantum gravity, which are each dubbed the emergent string and the decompactification limits, respectively \cite{Lee:2019wij}. The species scale defined in \eqref{SCD} was explicitly shown to satisfy these specific expectations in a diverse set of top-down examples \cite{vandeHeisteeg:2023dlw}. 

The above discussion can be illustrated through the instructive example of type IIA supergravity in 10 dimensions \cite{vandeHeisteeg:2023dlw,Calderon-Infante:2023uhz}, where the the species scale will depend on the dilaton $\phi$. The first non-zero higher-curvature correction to the supergravity action appears for the $\cR^4$ term where the coefficient $a_4$ is computed via the tree-level and one-loop four-graviton scattering amplitudes in type IIA and does not receive any further contributions \cite{Green:1997di}. Thus, we can identify the species scale with $a_4(\phi)^{-1/6}M_{\pl,10}$ 
\begin{equation}
    \Lambda_s=\frac{1}{(2\pi)^{1/8}} \left[\frac{3\zeta(3)}{\pi^2}e^{3\phi/\sqrt{2}}+e^{-\phi/\sqrt{2}}\right]^{-1/6}M_{\pl,10}\,.
\end{equation}
In this case, the theory has two infinite-distance limits, corresponding to $\phi\to\pm\infty$. On the one hand, when $\phi\to-\infty$, or equivalently $g_s\gg 1$, the species scale becomes
\begin{equation*}
    \Lambda_s\sim e^{\phi/6\sqrt{2}}M_{\pl,10}\propto M_{\pl,11}\,,
\end{equation*}
where $\left(M_{\pl,10}/M_{\pl,11}\right)^8=2\pi e^{-2\sqrt{2}\phi/3}$. This is precisely the limit at which the theory decompactifies to 11d M-theory. On the other hand, when $\phi\to+\infty$, or equivalently $g_s\ll 1$, the species scale becomes
\begin{equation*}
    \Lambda_s\sim e^{-\phi/2\sqrt{2}}M_{\pl,10}\propto M_s\,,
\end{equation*}
where $e^{-2\sqrt{2}\phi}M_{\pl,10}^8\propto M_{s}^8$. This is precisely the weak-coupling limit at which we have the emergent type IIA string. Therefore, this recovers the expected behavior in the infinite-distance limits of our moduli space.  Moreover, note that as expected $\Lambda_s(\phi)\lesssim M_{\pl,10}$ everywhere in the moduli space. 

\section{Phase transitions in quantum gravity}
\label{sec:Phase transitions}

Many of the unique features of quantum gravity can be traced back to the the existence and properties of black holes. In non-gravitational theories, the coupling does not have to be related to mass, i.e., couplings are not directly correlated with mass. Therefore, arbitrarily massive states can be weakly coupled (or even free) without any issues. However, in a gravitational theory, every state is coupled to the graviton with a coupling proportional to its mass. Therefore, beyond a mass threshold, the localized states are expected to become strongly coupled, resulting in black holes.

Luckily, we expect the effective theory of gravity including the Einstein term to be a good description of large sized black holes where curvatures are sufficiently small near the horizon.  However, as we decrease the size of a neutral black hole we expect the Hawking temperature to rise and beyond some temperature, the calculation of the free energy of the most stable saddle from the EFT becomes incorrect as the EFT does not detect the phase transitions of neutral black holes and also misses the perturbative instabilities associated to these black holes. Let us denote this limiting temperature by $T=\Lambda_{\rm BH}$, where the critical size black hole is $R_{\rm BH}\sim 1/\Lambda_{\rm BH}$.  Clearly $\Lambda_{\rm BH}\lesssim M_{\pl}$.  In fact, we would expect $\Lambda_{\rm BH}\lesssim \Lambda_s\lesssim M_{\pl}$ because the action becomes large near the horizon of black holes of size $1/\Lambda_s$, so the EFT surely breaks down at $\Lambda_s$ or below that scale.

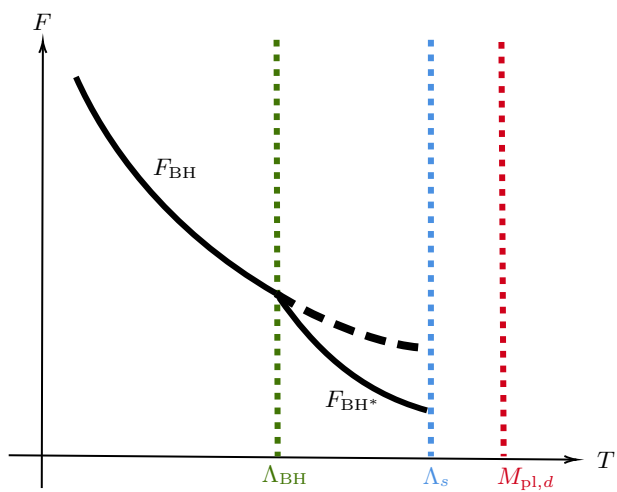
\begin{figure}[H]
\centering

\tikzset{every picture/.style={line width=0.75pt}} %set default line width to 0.75pt        

\begin{tikzpicture}[x=0.75pt,y=0.75pt,yscale=-0.55,xscale=0.55]
%uncomment if require: \path (0,464); %set diagram left start at 0, and has height of 464

%Straight Lines [id:da7246532960878103] 
\draw    (38,418) -- (551,421.98) ;
\draw [shift={(553,422)}, rotate = 180.45] [color={rgb, 255:red, 0; green, 0; blue, 0 }  ][line width=0.75]    (10.93,-3.29) .. controls (6.95,-1.4) and (3.31,-0.3) .. (0,0) .. controls (3.31,0.3) and (6.95,1.4) .. (10.93,3.29)   ;
%Straight Lines [id:da7111448080442837] 
\draw    (68,448) -- (69,41) ;
\draw [shift={(69,39)}, rotate = 90.14] [color={rgb, 255:red, 0; green, 0; blue, 0 }  ][line width=0.75]    (10.93,-3.29) .. controls (6.95,-1.4) and (3.31,-0.3) .. (0,0) .. controls (3.31,0.3) and (6.95,1.4) .. (10.93,3.29)   ;
%Straight Lines [id:da863966657616714] 
\draw [color={rgb, 255:red, 65; green, 117; blue, 5 }  ,draw opacity=1 ][line width=2.25]  [dash pattern={on 2.53pt off 3.02pt}]  (281,38) -- (282,419) ;
%Straight Lines [id:da6034636096683081] 
\draw [color={rgb, 255:red, 74; green, 144; blue, 226 }  ,draw opacity=1 ][line width=2.25]  [dash pattern={on 2.53pt off 3.02pt}]  (421,38) -- (421,419) ;
%Straight Lines [id:da2231031834799928] 
\draw [color={rgb, 255:red, 208; green, 2; blue, 27 }  ,draw opacity=1 ][line width=2.25]  [dash pattern={on 2.53pt off 3.02pt}]  (485,40) -- (487,419) ;
%Curve Lines [id:da3379940537474815] 
\draw [line width=2.25]    (99,72) .. controls (125,131) and (179.5,211) .. (280.5,270) ;
%Curve Lines [id:da002208040055933891] 
\draw [line width=3]  [dash pattern={on 7.88pt off 4.5pt}]  (280.5,270) .. controls (305.5,287) and (371.5,319) .. (419.5,320) ;
%Curve Lines [id:da7296190130789479] 
\draw [line width=2.25]    (283.5,274) .. controls (314,318) and (353.5,359) .. (417.5,377) ;

% Text Node
\draw (57,11.4) node [anchor=north west][inner sep=0.75pt]    {$F$};
% Text Node
\draw (569,412.4) node [anchor=north west][inner sep=0.75pt]    {$T$};
% Text Node
\draw (265,424.4) node [anchor=north west][inner sep=0.75pt]  [color={rgb, 255:red, 65; green, 117; blue, 5 }  ,opacity=1 ]  {$\Lambda _{\rm BH}$};
% Text Node
\draw (411,426.4) node [anchor=north west][inner sep=0.75pt]  [color={rgb, 255:red, 74; green, 144; blue, 226 }  ,opacity=1 ]  {$\Lambda _{s}$};
% Text Node
\draw (478,426.4) node [anchor=north west][inner sep=0.75pt]  [color={rgb, 255:red, 208; green, 2; blue, 27 }  ,opacity=1 ]  {$M_{\pl,d}$};
% Text Node
\draw (166,144.4) node [anchor=north west][inner sep=0.75pt]    {$F_{\text{BH}}$};
% Text Node
\draw (322,356.4) node [anchor=north west][inner sep=0.75pt]    {$F_{\text{BH}^{*}}$};

\end{tikzpicture}
    \caption{Free energy vs temperatures near important energy scales in $d$-dimensional quantum gravity. The black hole $(\mathrm{BH})$ transitions into a more stable object $(\mathrm{BH}^*)$ at temperature $\Lambda_\mathrm{BH}$. The thermodynamic description breaks down when the temperature reaches the species scale $\Lambda_s\lesssim M_{\pl,d}$.}
    \label{fig:intro to three scales}
\end{figure}

From the form of the effective action one may think that $\Lambda_{\rm BH}=\Lambda_s$ because one would naively expect that the only reason the effective action would not be describing the black hole correctly is that the higher-order terms become important.  As we will argue, this is not true and generically $\Lambda_{\rm BH}<\Lambda_s$ (Figure~\ref{fig:intro to three scales}).
The mechanism for this is that the nature of the black hole described by the IR properties of the EFT is not the stable one; at smaller radii (i.e., higher temperatures) there is another class of solutions which the EFT misses that have lower free energy.  So, the black holes undergo a transition at $T=\Lambda_{\rm BH}$ to these more stable states.  

We provide evidence for this in regions of weak-coupling in the QG landscape.  These are the asymptotic regions of parameter space where we expect to have either a tower of light KK modes or light fundamental strings \cite{Lee:2019wij}.  We now turn to these two cases and show that in both there is such a scale $\Lambda_{\rm BH}\lesssim \Lambda_s$ where the EFT incorrectly estimates the free energy of the more stable states to which the black holes with temperatures greater than $\Lambda_{\rm BH}$ transition.

We note that there can be more than one phase transition in the spectrum of massive states in quantum gravity. An example of this is when there are compact dimensions with parametrically different length scales.  In such cases we identify $\Lambda_{\rm BH}$ as the lowest such scale. The structure of these phase transitions are highly constrained from the top-down perspective and can be motivated using bottom-up arguments in the weak-coupling limits \cite{Bedroya:2024ubj}. In particular, the thermodynamics of weakly coupled particle towers at high temperatures has been studied in \cite{Cribiori:2023ffn,Basile:2023blg} using insights from black hole thermodynamics (see also \cite{Bedroya:2024ubj}).

\subsection{The Gregory--Laflamme transition}
\label{sec:GL transition}

In this section, we briefly review how the physics of black holes is affected by the presence of large extra dimensions. Suppose there are $d$ non-compact dimensions and $p$ large extra dimensions that have length scale $R$. The first question one could ask is, what is the meaning of a neutral black hole in $d$ dimensions from the perspective of the higher dimensional theory.

To answer this, let us start by considering the Schwarzschild black hole described by the following metric in $d$ dimensions
\begin{align}
    \dd s^2= -\left(1-\frac{r_{H}^{d-3}}{r^{d-3}}\right)\dd t^2&+\left(1-\frac{r_H^{d-3}}{r^{d-3}}\right)^{-1}\dd r^2\nonumber\\&+r^2\dd \Omega_{d-2}^2\,,
\end{align}
where $r_H$ is the horizon radius of the black hole. This solution is the dimensional reduction of the following metric in the higher dimensional theory.
\begin{align}
    \dd s^2= -&\left(1-\frac{r_{H}^{D-3-p}}{r^{D-3-p}}\right)dt^2+\sum_{i,j=1}^{p} g_{ij}(y)\dd y^i \dd y^j\nonumber\\
    &+\left(1-\frac{r_H^{D-3-p}}{r^{D-3-p}}\right)^{-1}\dd r^2+r^2\dd \Omega_{D-2-p}^2\,,
\end{align}
where $D=p+d$ and $g_{ij}(y)$ is the pullback metric on the compact manifold. The above solution is a black brane whose horizon is parametrized by $y^i$ as well as the angular coordinates. In other words, we have a black $p$-brane that is wrapped around the $p$ dimensional compact manifold. Therefore, a neutral black hole always lifts up to a higher-dimensional black brane that is uniformly wrapped around the extra dimensions. If the compact space has a symmetry (or approximate symmetry as in the case of large extra dimensions), any non-uniform black hole would involve a combination of KK charges.  Indeed, in the extreme limit when the black hole is localized in the extra dimensions, it would not be in a definite charge state, but a superposition thereof.  Therefore, it is natural to expect a neutral black hole to uplift to a uniformly wrapped black brane. This observation implies that black holes in some sense see the entirety of the compact space. This raises the question of can we probe the presence of extra dimensions using lower dimensional black holes. As is well known, and we will explain below, by lowering the mass of black holes, the black holes become unstable at $r_H\sim R$ due to the Gregory--Laflamme (GL) transition which provides a lower-dimensional window to probe the higher-dimensional theory. 
At this transition scale $r_H\sim R$, the uniformly wrapped black brane becomes unstable as was shown by Gregory and Laflamme \cite{Gregory:1993vy,Gross:1982cv,Reall:2001ag}. 

Even though, the perturbative analysis does not determine the final state, it is reasonable to expect that the black brane will break up into smaller pieces which will then clump up and collapse into black holes. In other words, the black hole will decompose into higher dimensional black holes that are localized in the extra dimensions. These black holes might subsequently join and form a larger higher-dimensional black hole.  Let us note an interesting feature of this transition. A black hole with vanishing KK momentum decays into a wavepacket of non-zero KK states. One might wonder if this process breaks the charge conservation when the compact space is symmetric. The answer is that it does not. It rather teaches us that neutral black holes for small enough black holes are superpositions of different charges and only on average have zero charge.  

Interestingly, the Gregory--Laflamme transition can also be understood as a thermodynamic transition. The mass scale at which the lower-dimensional black hole becomes perturbatively unstable, up to numerical constants, agrees with the mass scale where the entropies of the two solutions are equal. The entropy of the lower-dimensional black hole is given by 
\begin{align}
    S_d= \frac{4\pi^\frac{d+1}{2}}{\Gamma\left(\frac{d-1}{2}\right)}\left(\frac{M}{4\pi M_{\pl, d}}\right)^\frac{d-2}{d-3}\,,
\end{align}
while for the higher-dimensional black hole, it is given by 
\begin{align}
    S_D\simeq \frac{4\pi^\frac{D+1}{2}}{\Gamma(\frac{D-1}{2})}\left(\frac{M^{D-2}(2\pi R)^p}{(4\pi)^{D-2} M_{\pl, d}^{d-2}}\right)^\frac{1}{D-3}\,,
\end{align}
where $D=p+d$ is the total number of dimensions. Note that the factors of $R$ appear due to the relationship between the lower-dimensional and higher-dimensional reduced Planck mass
\begin{align*}
    M_{\pl, d}^{d-2}=M_{\pl, D}^{D-2}(2\pi R)^p\,,
\end{align*}
where, for simplicity, we have assumed the extra dimensions form a $p$-dimensional torus with radius $R$. The two expressions for entropies become equal at mass scale $M\sim R^{d-3}M_{\pl, d}^{d-2}$ and radius $\sim R$. For lower masses, the higher-dimensional black hole is more entropically favorable. 

We can also see the instability of the lower-dimensional black holes in the canonical ensemble where the temperature is kept fixed. The free energy of the $d$-dimensional and $D$-dimensional black holes in terms of the temperature $T$ are given by
\begin{align}
    F_d&=\frac{4\pi\left((d-3)\Gamma\left(\frac{d-1}{2}\right)\right)^{d-3}M_{\pl, d}^{d-2}}{\left(d-2\right)^{d-2}\left(\pi^\frac{d-1}{2}T\right)^{d-3}}\,, \\
    F_D&=\frac{4\pi\left((D-3)\Gamma\left(\frac{D-1}{2}\right)\right)^{D-3}M_{\pl, d}^{d-2}}{\left(D-2\right)^{D-2}\left(2\pi R\right)^p\left(\pi^\frac{D-1}{2}T\right)^{D-3}}\,.
\end{align}
The two expressions for free energies become equal at temperature $T\sim 1/R$. For higher temperatures, the higher-dimensional black hole has lower free energy and is therefore thermodynamically favorable.

It is interesting to note the consistency between the lower- and higher-dimensional black hole perspectives: one could ask, what is the maximum entropy one can obtain from fitting higher-dimensional black holes of mass $M$ and radius $r\propto (M/M_{\pl,D})^{1/{(D-3)}}M_{\pl,D}^{-1}\ll R$ inside the extra dimensions. We can fit $\sim (R/r)^p$ such black holes in $T^p$ and adding up their entropies comes out to 
\begin{align}
    S\sim\left(\frac{R}{r}\right)^p\left(rM_{\pl,D}\right)^{D-2}\sim (rM_{\pl, d})^{d-2}\,,
\end{align}
which is proportional to the entropy of a lower-dimensional black hole with the combined mass of $\sim \left(\frac{R}{r}\right)^p M$. In other words, the entropy formulas of lower and higher-dimensional black holes are non-trivially consistent with each other to ensure that the maximum entropy in a given region of the non-compact space follows the area law.

The presence of large extra dimensions leads to a tower of light KK states with masses proportional to $1/R$. What is remarkable about the GL transition is that it provides an important physical meaning to the KK energy scale $m_{\rm KK}\sim 1/R$. This is the temperature scale at which lower-dimensional black holes become unstable and undergo phase transition. In other words, the presence of large extra dimension is not only reflected in the spectrum of the theory at low masses ($m_{\rm KK}< M_{\pl,d}$) via KK states, but it is also observable in the UV states at much higher mass ($M=(RM_{\pl,d})^{d-3} M_{\pl,d}> M_{\pl,d}$) in the spectrum of black holes. 

Another interesting observation which we will discuss below, is that both manifestations of the extra dimensions (emergence of KK tower and instability of black holes) tell us that the low-energy EFT fails at describing black holes with temperatures $T\gg R^{-1}$ in a way that would not be observable to the EFT.  The basic idea is that the main effect of the large-radius compactification is to a good approximation simply reducing the higher-dimensional theory, that includes the higher-order terms, to the lower one.   To see this,  suppose in particular we are considering graviton scattering amplitudes in the lower-dimensional theory. If we were to compute this by including also the KK modes as intermediate states, this is already included as the momentum integral in the reduction of the higher-dimensional theory.  The only difference is replacing the continuous internal momenta $p$ with discrete sums of the form $n/R$ and we would expect this to be a small effect if the discretization is refined enough, i.e., if $1/R \ll M_{\pl,D}$.  In particular we expect the corrections to the amplitudes to be suppressed by powers of $1/(RM_{\pl,D})$ compared to the leading term given by suitable powers of $1/\Lambda_s(\phi_i)$. As explained in the Appendix~\ref{appendix}, the leading correction to the EFT and calculation of black hole free energy will be controlled by the species scale and is negligible for black holes with temperatures $T\ll\Lambda_s$. In summary, in the IR regime at temperatures $T \ll m_{\rm KK}$, the leading correction is generically sourced by the relevant and marginal terms descended from the higher-dimensional action rather than the KK particles running in the loops. On the other hand, in the energy regime where $T\gg m_{\rm KK}$, the collective effect of loop corrections is well-approximated by the higher-dimensional fields running in the loop. Therefore, the leading correction is still captured by the dimensional reduction of a higher-dimensional term.

In some sense, to see the new physics of KK states, one must have access to the KK states because their contribution to the scattering amplitudes with zero KK charge is only through loops and therefore is negligible. Naively, from the higher-derivative expansion of the EFT, one could have completely missed the KK states and thought that new physics only emerges at energy scale $\Lambda_s$. In fact, something similar happens for the black hole physics with an important exception. If one does not include the KK states in the lower-dimensional theory, there is no indication of a thermodynamic instability of the black holes and the existence of a more stable saddle (the higher-dimensional black hole). Nonetheless, one can experimentally observe this instability without having access to KK states, by looking at the thermodynamics of black holes with zero KK charge. Therefore, the low-energy EFT breaks down for describing the thermodynamics of black holes at temperatures  $T>m_{\rm KK}$. It is remarkable that contrary to particles with vanishing KK charge, black holes that carry no definite KK charge (neutral on the average) can see the presence of extra dimensions. 

\subsection{The black hole/string stars transition}
\label{sec:HP transition}

In this section we review how the black hole entropy formula fails at sufficiently high temperatures in weakly coupled string theories. 

The thermodynamics of a black hole in the canonical ensemble is described by a Euclidean solution with a thermal circle. The circumference of the thermal circle is $\beta$ in the asymptotic region, and it vanishes on the horizon \cite{Gibbons:1976ue}. The thermal circle has anti-periodic boundary condition for fermions which lowers the zero-point energy of winding strings. Consequently, the winding string becomes tachyonic when $\beta<\beta_{\rm H}$ for some $\beta_{\rm H}\sim M_s^{-1}$. The temperature $\beta_{\rm H}^{-1}\sim M_s$ is known as the Hagedorn temperature and is a manifestation of the exponential growth of the number of string excitations with respect to mass \cite{Hagedorn:1965st,Sathiapalan:1986db,Atick:1988si}. Naturally, one could ask whether the black hole entropy formula fails exactly at the Hagedorn temperature $T_{\rm H}$ or whether one must switch to a different description at a lower temperature?  

This was answered in \cite{Horowitz:1997jc} where Horowitz and Polchinski used the winding state in its non-tachyonic regime ($\beta>\beta_H$) to find new Euclidean saddles in spacetime dimensions $d<7$. One can use the Euclidean action to calculate the free energy and temperature of these solutions \cite{Chen:2021dsw}. There is a critical temperature $T_{\rm HP}<T_{\rm H}$ such that for all $T>T_{\rm HP}$,  the Horowitz-Polchinski (HP) solutions (also known as \textit{string stars}) have lower free energy and therefore are more stable. Therefore, there is a transition from the black holes to the more stable string stars for $T>T_{\rm HP}$. The transition temperature is of the same order, but strictly smaller, than the Hagedorn temperature.

Even though the HP solution does not exist in weakly coupled string theories in higher dimensions ($d\geq 7$), there is evidence that a strongly coupled solution replacing it might exist \cite{Balthazar:2022hno}.

Note that the light winding state plays a crucial role in the existence of the HP saddle in the Euclidean signature. The fact that the winding state around the thermal circle becomes massless at finite $\beta$ is the other side of the coin to the exponential growth of string tower. Therefore, the low-energy EFT without the string tower, would not be able to predict the HP saddle. In other words, the EFT calculation of the free energy of these more thermodynamically stable saddles become incorrect due to it not having the tower of excited string states at $T_{\rm HP}$.

It is interesting to note a parallel between GL transition and the HP transition. In both phase transitions, the topology of the Euclidean saddle changes. The Euclidean black hole has a topology of $\mathbb{R}^2\times S^{d-2}$ while the HP solution has a topology of $\mathbb{R}^{d-1}\times S^1$. The main difference comes from the fact that the thermal circle shrinks in the black hole case but does not shrink to zero size in the HP solution. Using this topology change, the authors in \cite{Chen:2021dsw} showed that there is a supersymmetric obstruction for a smooth transition between black holes and HP solutions as target spaces of type II strings. One of the topological indices used was the Witten index which measures the Euler characteristic of the target space \cite{Witten:1982df}. 
We note that the same argument can be applied to the Gregory--Laflamme transition, as the corresponding black hole solutions in $d$ and $D$ dimensions have different Euler characteristics\footnote{Consider the spacetime with $n$ non-compact dimensions and a compact space $T^{m}$. Suppose $L_{n,m}$ is the space resulting from capping off the Euclidean black hole that is localized in compact space with $B^2\times T^{m}\times S^{n-2}$. Consider two parallel hyperplanes that are seperated in a non-compact direction and contain the black hole in between them. We divide $L_{m,n}$ into three regions across these hyperplanes. Let us identify the boundaries of the middle part with each other and identify the boundaries of the other two regions with each other. This process transforms $L_{m,n}$ into $L_{m+1,n-1}\sqcup S^{n}\times T^{m}$. Using the additivity of the Euler characteristic and $\chi(S^{n}\times T^m)=1+(-1)^n$, by induction we find
\begin{align}
    \chi(L_{p,d})=2\left(1+(-1)^{d+p}\right)-\sum_{i=0}^{p-1}\left(1+(-1)^{d+p-i}\right)\,.
\end{align}
The above Euler characteristic can never be equal to the Euler characteristic corresponding to the lower dimensional black hole $\chi(L_{0,d})=2(1+(-1)^d)$. Therefore, there is a topological obstruction for smoothly connecting the world sheet theories of type II strings in the backgrounds of lower and higher dimensional black holes in a non-singular way, as in the case of HP.}.

\section{A new swampland conjecture}
\label{sec:Swampland conjecture}

The above discussion in section~\ref{sec:Phase transitions} suggests that in addition to $M_{\pl}$ and $\Lambda_s$ in all the weak coupling regimes of $\cM_{\rm QG}$, there exists another scale $\Lambda_{\rm BH}\lesssim\Lambda_s\lesssim M_{\pl}$ which marks the smallest temperature, or the inverse radius of the largest black hole at which the EFT misses the most stable black hole and a phase transition to a more stable state occurs.   In both cases the EFT of Einstein action (and higher-order corrections) misses these solutions as it does not incorporate the infinitely many light elements of the tower (the KK tower in the GL case and the light tower of strings manifested by Hagedorn behavior/thermal winding string condensate for the string tower in the HP case). 

In particular, in the emergent string limit where $\Lambda_s$ is identified with the Hagedorn temperature $\Lambda_s=T_{\rm H}$, the black hole description breaks down at the Horowitz--Polchinski temperature $\Lambda_{\rm BH}=T_{\rm HP}$ where the black holes transition into self-gravitating strings. In the decompactification limit where the $d$-dimensional quantum gravitational theory decompactifies to a $D$-dimensional theory, $\Lambda_s$ is identified as $M_{\pl,D}$ while the $d$-dimensional black hole description breaks down at $\Lambda_{\rm BH}\sim m_{\rm KK}$ scale where the $d$-dimensional black hole transitions into a $D$-dimensional black hole. In both of these limits, the transition occurs at energies below the species scale, i.e., $T_{\rm HP}\lesssim T_{\rm H}$ and $m_{\rm KK}\lesssim M_{\pl,D}$ or equivalently in both cases $\Lambda_{\rm BH}\lesssim \Lambda_s$. Different from the species scale, neither $T_{\rm HP}$ nor $m_{\rm KK}$ is naturally encoded within the action of the EFT. Instead, it comes purely from studying phase transitions of black holes. In other words, this scale arises from the non-analyticity of black hole entropy and free energy.  The black hole transitions into a more stable physical state with lower free energy at temperature $\Lambda_{\rm BH}$.

Here we have mainly concentrated on the weak coupling regime where we have seen $\Lambda_{\rm BH}$ is similar to the mass scale of the light tower.  However their ratio cannot be moduli independent.  For example consider the large extra dimension case. In this case, with enough supersymmetry, the mass of the KK tower $1/R$ is not corrected as we vary $R$. However, there are corrections to the Einstein action which will affect the point at which the GL transition happens, and thus the ratio of $\Lambda_{\rm BH}/m_{\rm KK}$ is expected to depend on $R$.   For example, for M-theory on $S^1$, the $\cR^4$ corrections \cite{Green:1997di}  correct the entropy of the lower and the higher dimensional black holes \cite{Chen:2021qrz} and the location where they meet which predicts the GL transition point changes. In this case, the 11d black hole has entropy 
\begin{equation*}
    S_{\rm M}^{\rm (11d)}\sim (M_{\pl,10}r)^8\left(\frac{r}{R}\right)\left[1-\frac{132609}{4}\frac{1}{(M_{\pl,11}r)^6}\right]\,,
\end{equation*}
whereas the 10d black hole in type IIA supergravity theory has entropy
\begin{equation*}
    S_{\rm IIA}^{\rm (10d)}\sim (M_{\pl,10}r)^8\left[1-\frac{80420}{3}\frac{a_4(\phi)}{(M_{\pl,10}r)^6}\right]\,,
\end{equation*}
where $a_4(\phi)$ is as defined earlier. A rough estimate of $\Lambda_{\rm BH}$ in the decompactification limit of the type IIA theory can be found by solving for $r$ such that $S_{\rm IIA}^{(\rm 10d)}=S_{\rm M}^{(\rm 11d)}$. Hence, in the large-volume limit, we obtain
\begin{equation}
\label{eq:LambdaBH IIA}
    \frac{\Lambda_{\rm BH}}{m_{\rm KK}}\propto 1+a\cdot e^{-4\sqrt{2}\phi}+\dots\,,
\end{equation}
where $a\approx 269.4$.
Therefore, the ratio $\Lambda_{\rm BH}/m_{\rm KK}$ becomes $\phi$-dependent at large $\phi$. Thus, $\Lambda_{\rm BH}$ {\it cannot} be identified as the mass scale of the lightest state for all moduli.

Note that the above observations hold true in both the weak-coupling regime of the fundamental string and the large-volume regime of compactifications. Consequentially, the notion of this phase transition does not strictly live in the infinite-distance limits of $\cM_{\rm QG}$, and $\Lambda_{\rm BH}$ is expected to be well-defined across the moduli space. Thus, $\Lambda_{\rm BH}$ should be globally defined in $\cM_{\rm QG}$ as the temperature at which the Schwarzschild black hole becomes unstable due to the existence of a more stable saddle whose free energy is not seen by the EFT. Therefore, all of this motivates us to propose the following Swampland conjecture:

    {\it Any consistent EFT description of $d$-dimensional quantum gravity must exhibit three scales across its moduli space $\cM_{\rm QG}$: 1) the $d$-dimensional Planck scale, $M_{\pl,d}$ which controls the strength of the Einstein term; 2) the species scale, $\Lambda_s$, where the higher-order gravitational corrections become important; and 3) the black hole scale, $\Lambda_{\rm BH}$, where at this temperature, the black hole predicted by EFT undergoes a phase transition to a more stable solution. Furthermore, $\Lambda_{\rm BH}\lesssim \Lambda_s\lesssim M_{\pl,d}$ everywhere in $\cM_{\rm QG}$ and $\Lambda_{\rm BH}$ approaches the mass scale of the lightest tower at large distances in field space.}

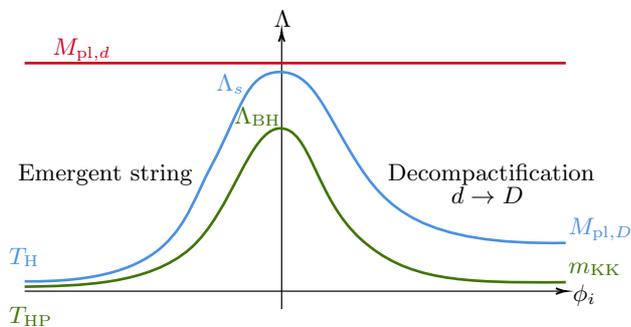
\begin{figure}[H]
\centering

\tikzset{every picture/.style={line width=0.45pt}} 

\begin{tikzpicture}[x=0.75pt,y=0.75pt,yscale=-0.45,xscale=0.45]
\draw    (30,320) -- (630,320) ;
\draw [shift={(630,320)}, rotate = 180.1] [color={rgb, 255:red, 0; green, 0; blue, 0 }  ][line width=0.75]    (10.93,-3.29) .. controls (6.95,-1.4) and (3.31,-0.3) .. (0,0) .. controls (3.31,0.3) and (6.95,1.4) .. (10.93,3.29)   ;
\draw  (315,340) -- (315,30) ;
\draw [shift={(315,30)}, rotate = 89.82] [color={rgb, 255:red, 0; green, 0; blue, 0 }  ][line width=0.75]    (10.93,-3.29) .. controls (6.95,-1.4) and (3.31,-0.3) .. (0,0) .. controls (3.31,0.3) and (6.95,1.4) .. (10.93,3.29)   ;
\draw [color={rgb, 255:red, 74; green, 144; blue, 226 }  ,draw opacity=1 ] [line width=1]  (30,309) .. controls (199,311) and (204,242) .. (237,180) .. controls (270,118) and (275,76) .. (312,75) .. controls (349,74) and (362,117) .. (396,172) .. controls (430,227) and (469.5,269) .. (630,266) ;
\draw  [color={rgb, 255:red, 65; green, 117; blue, 5 }  ,draw opacity=1 ] [line width=1] (30,315) .. controls (91,314) and (168,313) .. (214,272) .. controls (260,231) and (279,139) .. (313,138) .. controls (347,137) and (352,216) .. (403,262) .. controls (454,308) and (511,311) .. (630,310) ;
\draw  [color={rgb, 255:red, 208; green, 2; blue, 27 }  ,draw opacity=1 ][line width=1] (30,65) -- (630,65) ;

\draw (304,3) node [anchor=north west][inner sep=0.75pt]    {$\Lambda $};
\draw (635,310) node [anchor=north west][inner sep=0.75pt]    {$\phi_i$};
\draw (240,75) node [anchor=north west][inner sep=0.75pt]  [color={rgb, 255:red, 74; green, 144; blue, 226 }  ,opacity=1 ]  {$\Lambda _{s}$};
\draw (260,111) node [anchor=north west][inner sep=0.75pt]  [color={rgb, 255:red, 65; green, 117; blue, 5 }  ,opacity=1 ]  {$\Lambda _{\rm BH}$};
\draw (10,268.4) node [anchor=north west][inner sep=0.75pt]  [color={rgb, 255:red, 74; green, 144; blue, 226 }  ,opacity=1 ]  {$T_{\rm H}$};
\draw (10,335) node [anchor=north west][inner sep=0.75pt]  [color={rgb, 255:red, 65; green, 117; blue, 5 }  ,draw opacity=1 ]  {$T_{\rm HP}$};
\draw (630,285) node [anchor=north west][inner sep=0.75pt] [color={rgb, 255:red, 65; green, 117; blue, 5 }  ,draw opacity=1 ] {$m_{\rm KK}$};
\draw (630,235) node [anchor=north west][inner sep=0.75pt]  [color={rgb, 255:red, 74; green, 144; blue, 226 }  ,opacity=1 ]  {$M_{\pl,D}$};
\draw (20,175) node [anchor=north west][inner sep=0.75pt]   [align=left] {Emergent string};
\draw (430,175) node [anchor=north west][inner sep=0.75pt]   [align=left] {Decompactification};
\draw (500,200) node [anchor=north west][inner sep=0.75pt]   [align=left] {$d\to D$};
\draw (58,35) node [anchor=north west][inner sep=0.75pt]  [color={rgb, 255:red, 208; green, 2; blue, 27 }  ,opacity=1 ]  {$M_{\pl,d}$};

\end{tikzpicture}
\caption{A schematic illustration of the relation among $M_{\pl,d}$, $\Lambda_s$, and $\Lambda_{\rm BH}$ for a $d$-dimensional quantum gravitational theory with respect to the moduli parameter $\phi_i$ according to our conjecture. In particular, in the asymptotic regions of moduli space, the expected behavior from the GL and the HP transitions are shown.}
\label{fig:Conjecture}
\end{figure}

It is amusing to consider the case of type IIA in 10 dimensions for which the moduli space is parameterized by the string coupling (or equivalently the radius of the extra circle).  In this case we would expect a behavior shown in Figure \ref{fig:Conjecture}.  It is interesting to note that in this case we expect the GL transition to be connected continuously to a transition analogous to the HP transition.\footnote{A similar, but distinct, story unfolds in $(1+1)$d maximally supersymmetric Yang--Mills theory compactified on a generic tori, $T_{\tau}^2$, with appropriate boundary conditions imposed on the fermions \cite{Aharony:2005ew}. In particular, by varying the complex parameter $\tau$, the analog of the GL transition in gauge theory can then be continuously mapped to the confinement/deconfinement transition.}

One may ask whether one can identify $\Lambda_{\rm BH}\sim m_{\rm lightest}$ everywhere in moduli space and not just asymptotically where $m$ is the lightest mass excitation not included in the EFT.  This cannot be true everywhere.  For example if we have a point in the moduli space where some conformal field theory appears, then we have $m_{\rm lightest} \rightarrow 0$. However, since this tower is not weakly coupled, it does not make either the species scale $\Lambda_s$ or $\Lambda_{\rm BH}$ vanish since we are just adding finitely many weakly-coupled degrees of freedom to the theory which should not dramatically impact black hole behaviour.  One could instead hope that $m_\text{lightest}$ is to be identified with the mass of the lightest tower that becomes light and weakly-coupled in some asymptotic regime of the moduli space.   However, as we just explained, the ratio $\Lambda_{\rm BH}/m_\text{lightest}$ receives moduli dependent corrections as we move away from the infinite distance limits.

Unlike the $\Lambda_s$ where analytic computations are possible in some cases \cite{vandeHeisteeg:2022btw,vandeHeisteeg:2023dlw}, it seems that $\Lambda_{\rm BH}$ is much harder to compute explicitly except in the infinite distance limits.  
Using the fact that asymptotically $\Lambda_{\rm BH}\sim m_{\rm lightest}$, it implies from the work \cite{Castellano:2023stg,Castellano:2023jjt} that at least asymptotically

$$\nabla\ln\Lambda_s \cdot \nabla \ln\Lambda_{\rm BH}=\frac{1}{d-2}\,.$$

Since we have a well defined notion of $\Lambda_{\rm BH}$ and $\Lambda_s$ everywhere in moduli space, it is natural to ask if this relation can hold in the interior of the moduli space as well.\footnote{Extending the equality of \cite{Castellano:2023jjt,Castellano:2023stg} into the interior of moduli space has been attempted in \cite{Rudelius:2023spc} with specially chosen BPS scales for 5d $\cN=1$ theories. However, these (Dirac-paired) BPS scales are distinct from the scales of interest in this present paper.}  If this were true, since $\Lambda_s$ does have critical points in the interior of moduli space, it implies that at such points $\Lambda_{\rm BH}$ vanishes or $|\nabla \Lambda_{\rm BH}|$ blows up. As we argued above, we do not expect  $\Lambda_{\rm BH}$ to vanish in the interior of the moduli space where the number of weakly coupled particles is finite. Similarly, we expect the contribution of finite number of weakly coupled particles to $\nabla\Lambda_{\rm BH}$ to be regular (and moreover not to lead to divergent $\Lambda_{\rm BH}$ as this would). This means that the above relation cannot hold everywhere in the moduli space. Indeed, as we will explicitly check below, in the case of type IIA in 10 dimensions at large coupling, this equality does not hold at the subleading level.

In \cite{vandeHeisteeg:2023dlw,vandeHeisteeg:2023ubh} evidence was provided for the bound
$$\nabla \ln\Lambda_s \cdot \nabla \ln\Lambda_{\rm s}\lesssim \frac{1}{d-2}\,.$$
This naturally suggests that perhaps the correct relation involving both $\Lambda_s$ and $\Lambda_{\rm BH}$ is also an inequality, instead of the equality which holds only asymptotically, namely
\begin{equation}
\nabla \ln\Lambda_s \cdot \nabla \ln\Lambda_{\rm BH}\lesssim \frac{1}{d-2}\,.
\end{equation}

The existence of critical points for $\Lambda_s$ would be compatible with this. In particular, this relation,  
in the large-volume regime of the type IIA theory according to the perturbative corrections given in \eqref{eq:LambdaBH IIA}, can be computed as
\begin{align}
    \nabla\ln\Lambda_s\cdot\nabla\ln\Lambda_{\rm BH}&=\frac{1}{8}\left(1-e^{-2\sqrt{2}\phi}+\mathcal{O}(e^{-4\sqrt{2}\phi})\right)\nonumber\\&\leq \frac{1}{8}=\frac{1}{d-2}\,.
\end{align}
providing an example of this conjecture. It would be important to develop effective tools to compute $\Lambda_{\rm BH}$ in the interior of moduli space and to investigate the validity of this inequality.

One may ask, what are these three quantum gravity scales in our universe.  For the Dark Dimension scenario, motivated by the Swampland program,  these three scales end up being vastly different (see \cite{Vafa:2024fpx} for a review) and are given by 
\begin{align}
    &\Lambda_{\rm BH}\sim .01\text{eV},~~\Lambda_s\sim10^{9}\text{GeV}\,,~~M_{\pl}\sim2.4\times 10^{18}\text{GeV}\nonumber\,,
\end{align}
which in Planck units are related to the Dark energy $\Lambda\sim 10^{-122}$ by
\begin{align}
    &\Lambda_{\rm BH}\sim \Lambda^{3/12},~~\Lambda_s\sim \Lambda^{1/12},~~M_{\pl}\sim\Lambda^0 \nonumber\,,
\end{align}
(and the missing power of $\Lambda$ above is close to the weak scale $\sim \Lambda^{2/12}$ which is also partially explained in the Dark Dimension scenario).

\subsubsection*{Acknowledgments} 
We would like to thank Ruth Gregory, Juan Maldacena, Shiraz Minwalla, Rashmish Mishra, and Max Wiesner for valuable discussions. We would also like to thank Alberto Castellano and Dieter L\"ust for valuable comments on earlier versions of this letter. This work is supported in part by a grant from the Simons Foundation (602883,CV), the DellaPietra Foundation, and by the NSF grant PHY-2013858. AB is supported in part by the Simons Foundation grant number 654561 and by the Princeton Gravity Initiative at Princeton University.

\bibliography{papers}
~
\clearpage
\appendix

\section{Running of EFT coefficients}
\label{appendix}

In this section, we show that the dominant sub-leading contributions to the Bekenstein--Hawking entropy of a black hole, arising from higher-curvature terms appearing in the action, is always given by powers of the species scale. We will also review the energy dependence of a generic higher-derivative corrections to the Einstein action. 

In the EFT approach, we keep a finite number of terms which comes at the cost of omitting some of the information of the scattering amplitudes. The lost information is encoded in the energy dependence of the coefficients of the EFT. To study the corrections to the Bekenstein--Hawking entropy of a black hole, we can use the gravitational action evaluated at a energy scale $\mu$ set by the inverse radius of the given black hole.

We start by reviewing how particles running in loops can generate gravitational higher-derivative corrections. We explain that KK particles generate loop corrections $a^{\rm KK}_{n}(\mu)\mathcal{O}_n(\mathcal{R}^n)$ at energy scale $\mu$ such that 
\begin{align}
\begin{split}
\label{KKC}
    &\mu\ll m_{\rm KK}: a^{\rm KK}_{n}\sim \frac{1}{\mu^{2n-d}}\,,\\
    &\mu\gg m_{\rm KK}: a^{\rm KK}_{n}\sim \frac{M_{\pl,d}^{d-2}}{\mu^{2n-D}M_{\pl,D}^{D-2}}\,,
\end{split}
\end{align}
where $M_{\pl,D}$ is the higher-dimensional Planck mass in $D$ dimensions.

If the particle is massless (e.g., gravitons) the particle can generate a series of irrelevant operators by running in the loop. At energy scale $\mu$, the $n$-graviton amplitude with a massless loop contributes $\sim M_{\pl,d}^{-n(d-2)/2}\mu^{d}$ to the amplitude which results in corrections of the form $\mu^{d-2n}\mathcal{O}_n(\mathcal{R})$. The divergence of the coefficient in the $\mu\to 0$ limit is also known as the massless threshold behavior. On the other hand, a massive particle, running in a loop, with mass $m\gg \mu$ contributes $\sim m^{d-2n}$ to $a^{\rm KK}_{n}$. However, if the particle is very light, i.e., $m\ll \mu$, the particle resembles a massless mode and its contribution to $a^{\rm KK}_{n}$ is again $\sim \mu^{d-2n}$. Now, suppose we have a tower of KK modes resulting from compactifying a $D$-dimensional theory down to $d$ dimensions. At energy scale $\mu\ll m_{\rm KK}$, the contribution of the KK tower to the coefficient of $\mathcal{O}_n(\mathcal{R})$ behaves as $\mathcal{O}(\mu^{d-2n})+\mathcal{O}(m_{\rm KK}^{d-2n})$ where the first term is dominant and comes from the massless particle running in a loop. However, at energies $\mu\gg m_{\rm KK}$, there are $\sim (\mu/m_{\rm KK})^{D-d}$ light particles. Therefore, the loop corrections go as $\mathcal{O}((\mu/m_{\rm KK})^{D-d}\mu^{d-2n})+\mathcal{O}(m_{\rm KK}^{d-2n})$ where the first term comes from light states with $m\ll\mu$ and the second term from the heavy states with $m\gg \mu$. Note that in this case we can use the identity $m_{\rm KK}^{D-d}\sim M_{\pl,D}^{D-2}M_{\pl,d}^{2-d}$ to express the first term as $\mathcal{O}(\frac{M_{\pl,d}^{d-2}}{\mu^{2n-D}M_{\pl,D}^{D-2}})$. In both cases $\mu\ll m_{\rm KK}$ and $\mu\gg m_{\rm KK}$, the dominant term is the first term which explains the order of magnitudes expressed in \eqref{KKC} for $a_n^{\rm KK}$. Note that the dominant term is only a function of $\mu$ and the higher-dimensional Planck mass, and not the KK mass. 

Now, let us study the effect of loop corrections \eqref{KKC} to the entropy of $d$-dimensional Schwarzschild black holes. The Bekenstein--Hawking formula results from the Einstein term. Therefore, to compare the effect of the correction to the Bekenstein--Hawking entropy, we compare $M_{\pl,d}^{d-2}\mathcal{R}$ with $a_n(\mu)\mathcal{R}^n$, at energy scale $\mu\sim \sqrt{R}$. Using \eqref{KKC}, one can see that as long as $\mu\ll M_{\pl,D}$, the correction is sub-leading. Therefore, we conclude that for any energy scale below the species scale, the EFT corrections are sub-leading. 

Let us highlight four points about the energy dependence of the coefficient $a_n(\mu)$. 
\begin{itemize}
    \item The leading correction to the black hole entropy as suggested by the EFT depends on the species scale and not the KK scale.
    \item In the decompactification limit ($\mu$ fixed while $m_{\rm KK}\to 0$), the coefficient $a^{\rm KK}_{n}\sim M_{\pl,d}^{d-2}\mu^{D-2n}M_{\pl,D}^{2-D}$ can be thought of as the dimensional reduction of the term $\sim \mu^{D-2n}\mathcal{R}^{n}$ in $D$ dimensions which is simply the $D$-dimensional massless threshold. In other words, integrating out the KK loops combined with the lower-dimensional massless threshold reproduces the higher-dimensional massless threshold. This is a well-known fact that has been extensively tested in string theory, see e.g., \cite{Green:1999pu}.
    \item We note that our definition of the EFT differs from some literature, e.g., \cite{Green:2010wi}, in which the massless threshold, also known as the non-analytic part of the coefficient, is excluded from $a_n$. The EFT, defined in that way, will have coefficients that scale as $\tilde a_n\sim m_{\rm KK}^{d-2n}$ at energies $\mu\ll m_{\rm KK}$. This separation is sometimes done to avoid a diverging EFT coefficient at low energies. However, such separation will not be necessary for us, since as explained before, the combination $a_{n}(\mu)\mathcal{O}_n(\mathcal{R})$ for black holes with curvature scale $\mu$ will be sub-leading, even if $a_n(\mu)$ is large.
    \item The coefficient $a_n(\mu)$ at energy scale $\lambda_s\sim M_{\pl,D}$ becomes $a^{\rm KK}_{n}\sim M_{\pl,d}^{d-2}M_{\pl,D}^{2-2n}$, which is the expected tree-level correction. Therefore, the coefficients of EFT are controlled by the species scale, when they are evaluated at the species scale, and beyond that energy scale the EFT breaks down.
\end{itemize}

\end{document}